\documentclass[11pt,a4paper]{article}
\usepackage[T1]{fontenc}
\usepackage{setspace}
\usepackage[latin1]{inputenc}
\usepackage{amsmath}
\usepackage{amssymb}
\usepackage{setspace}
\usepackage{esint}

\date{}

\allowdisplaybreaks

\numberwithin{equation}{section}

\title{Notes on the pure spinor $b$ ghost}

\author{Renann Lipinski Jusinskas\thanks{renannlj@ift.unesp.br%
}}

\begin{document}

\maketitle
\begin{center}
ICTP South American Institute for Fundamental Research \\
Instituto de F\'isica Te\'orica, UNESP - Univ. Estadual Paulista \\
Rua Dr. Bento T. Ferraz 271, 01140-070, S\~ao Paulo, SP, Brasil.
\par\end{center}

\

\begin{abstract}In this work, a particular BRST-exact class of deformations of the $b$ ghost in the non-minimal pure spinor formalism is investigated and the impact of this construction in the $\mathcal{N}=2$ $\hat{c}=3$ topological string algebra is analysed. As an example, a subclass of deformations is explicitly shown, where the $U\left(1\right)$ current appears in a conventional form, involving only the ghost number currents. Furthermore, a $c$ ghost like composite field is introduced, but with an unusual construction.\end{abstract}

\section{Introduction}

\

The pure spinor (PS) formalism \cite{Berkovits:2000fe} is an \emph{ad
hoc} approach to the quantization of the superstring, in the sense
that the gauge fixing procedure that provides the BRST-like description
has not yet been discovered. It allows explicit Lorentz covariant
computations in the elegant language of $D=10$ superfields, gathering
together the advantages of the other main formalisms (RNS and Green-Schwarz)
without most of their restrictions. The Green-Schwarz formulation
\cite{Green:1983wt} does not allow a covariant split between first
and second class constraints and quantization is achieved in the (semi)
light cone gauge, with the introduction of the interaction-point operators.
As for the Ramond-Neveu-Schwarz (RNS) string, amplitude computations
require the sum over spin structures (implied by world-sheet supersymmetry
and related to GSO projection), integration over super \emph{moduli}
space and the introduction of picture-changing and spin operators
\cite{Friedan:1985ge}, lacking explicit space-time supersymmetry
(the algebra closes up to a picture changing operation) and making
the Ramond sector hard to deal with.

The world-sheet origin of the PS formalism, however, is still unknown,
as reparametrization symmetry is hidden. Therefore, understanding
the properties of the $b$ ghost is a fundamental task in providing
a better understanding of the formalism itself, its foundations and
potential developments.

Since its picture-raised introduction in \cite{Berkovits:2004px},
and later extension with the inclusion of the non-minimal variables
\cite{Berkovits:2005bt}, the $b$ ghost has been successfully used
in computing loop amplitudes, due to the critical topological string
interpretation of the formalism. However, the general properties of
$b$ are non-trivial, as it is also non-trivially composed. Its rich
structure has been explored over the years \cite{Berkovits:2006vi,Oda:2007ak,Chandia:2010ix,Jusinskas:2013yca,Berkovits:2013pla},
but it is not yet completely understood.

This work discusses two topics. The first one is the non-uniqueness
of $b$. As long as it is defined to satisfy $\left\{ Q_{0},b\right\} =T$,
where $Q_{0}$ is BRST charge and $T$ is the energy-momentum tensor,
any other $b'=b+\delta b$ with $\left\{ Q_{0},\delta b\right\} =0$
satisfies the same relation, being as good as the ``original'' one.
Since the BRST cohomology is nontrivial only for world-sheet scalars,
$b'$ and $b$ can only differ by a BRST-exact term and loop computations
using different versions of the $b$ ghost will always give the same
result. The general form of the deformations $\delta b$ will be constrained
in order to show explicitly the invariance of the $\mathcal{N}=2$
$\hat{c}=3$ algebra, including nilpotency of the deformed $b$ ghost.
The second topic is a $c$ ghost like field, motivated by a naive
analogy with the other formalisms, where the $\left(b,c\right)$ system
arises in gauge fixing the reparametrization symmetry. The pure spinor
formalism does not have a natural conformal weight $-1$ field to
act as the conjugate of $b$ and its existence is intriguing. A $c$
ghost like field will be introduced, but with an unusual construction,
and it will be shown to satisfy the expected properties of such object.

The paper is organized as follows. Section \ref{sec:review} is a
short review of the pure spinor formalism. Section \ref{sec:deformations}
discusses a particular BRST-exact class of deformations of the $b$
ghost and how it impacts the $\mathcal{N}=2$ algebra, including a
specific example. Finally, section \ref{sec:cghost} introduces the
$c$ ghost. The appendix contains the conventions and the full set
of fundamental OPE's that are being used in this work.

\section{Review of the non minimal pure spinor formalism\label{sec:review}}

\

Starting with the Green-Schwarz-Siegel action (holomorphic sector),
\begin{equation}
S_{\textrm{matter}}=\frac{1}{2\pi}\int d^{2}z\left(\frac{1}{\alpha'}\partial X^{m}\overline{\partial}X_{m}+p_{\beta}\overline{\partial}\theta^{\beta}\right),\label{eq:matteraction}
\end{equation}
the pure spinor formalism introduces a bosonic spinor conjugate pair
$\left(\lambda^{\alpha},\omega_{\alpha}\right)$, with action
\begin{equation}
S_{\lambda}=\frac{1}{2\pi}\int d^{2}z\left(\omega_{\alpha}\bar{\partial}\lambda^{\alpha}\right).\label{eq:actionminps}
\end{equation}

The BRST charge is, then, defined to be
\begin{equation}
Q\equiv\oint dz\left(\lambda^{\alpha}d_{\alpha}\right).\label{eq:BRSTmin}
\end{equation}
Here,
\begin{equation}
d_{\alpha}=p_{\alpha}-\frac{1}{\alpha'}\partial X^{m}\left(\theta\gamma_{m}\right)_{\alpha}-\frac{1}{4\alpha'}\left(\theta\gamma^{m}\partial\theta\right)\left(\theta\gamma_{m}\right)_{\alpha},\label{eq:susyderivative}
\end{equation}
the usual free field construction of the Green-Schwarz fermionic constraints.
The supersymmetric momentum is defined to be:
\begin{equation}
\Pi^{m}=\partial X^{m}+\frac{1}{2}\left(\theta\gamma^{m}\partial\theta\right).\label{eq:susymomentum}
\end{equation}
Note that
\begin{equation}
\left\{ Q,Q\right\} =-\frac{2}{\alpha'}\oint dz\left(\lambda\gamma^{m}\lambda\right)\Pi_{m},
\end{equation}
and nilpotency is achieved when $\left(\lambda\gamma^{m}\lambda\right)=0$,
the $D=10$ pure spinor condition. As discussed in \cite{Berkovits:2000fe},
the formalism successfully quantize the superstring theory, having
the advantage of being manifestly supersymmetric and Lorentz covariant.

The non-minimal version of the pure spinor formalism includes a new
set of variables, $\left(\overline{\lambda}_{\alpha},r_{\alpha}\right)$,
and their conjugates $\left(\overline{\omega}^{\alpha},s^{\alpha}\right)$.
The corresponding action is
\begin{equation}
S_{\overline{\lambda}}=\frac{1}{2\pi}\int d^{2}z\left(\bar{\omega}^{\alpha}\bar{\partial}\bar{\lambda}_{\alpha}+s^{\alpha}\bar{\partial}r_{\alpha}\right),\label{eq:actionnmps}
\end{equation}
and the non-minimal BRST current is defined to be
\begin{equation}
J_{BRST}\equiv\lambda^{\alpha}d_{\alpha}+\overline{\omega}^{\alpha}r_{\alpha}.\label{eq:BRSTnonmin}
\end{equation}
For later convenience, the Laurent modes of the BRST current will
be explicitly defined by
\begin{equation}
Q_{n}\equiv\oint dzz^{n}J_{BRST}\left(z\right).\label{eq:BRSTmodes}
\end{equation}
Through the quartet argument, it can be shown that $Q_{0}$, the non-minimal
BRST charge, has the same cohomology of \eqref{eq:BRSTmin}.

The bosonic ghost $\overline{\lambda}_{\alpha}$ is also a pure spinor,
whereas the fermionic variable $r_{\alpha}$ satisfy a consistency
condition, since
\begin{eqnarray}
\left[Q_{0},\overline{\lambda}\gamma^{m}\overline{\lambda}\right] & = & -2\overline{\lambda}\gamma^{m}r.
\end{eqnarray}
Summarizing, $\lambda$, $\overline{\lambda}$ and $r$ are constrained
through
\begin{equation}
\left(\lambda\gamma^{m}\lambda\right)=\left(\overline{\lambda}\gamma^{m}\overline{\lambda}\right)=\left(\overline{\lambda}\gamma^{m}r\right)=0,\label{eq:psconstraints}
\end{equation}
implying that each one of them has only $11$ independent components,
and that the actions \eqref{eq:actionminps} and \eqref{eq:actionnmps}
are gauge invariant by\begin{subequations}
\begin{eqnarray}
\delta_{\epsilon}\omega & = & \epsilon^{m}\left(\gamma_{m}\lambda\right)\\
\delta_{\overline{\epsilon},\phi}\bar{\omega} & = & \overline{\epsilon}^{m}\left(\gamma_{m}\bar{\lambda}\right)+\phi^{m}\left(\gamma_{m}r\right),\\
\delta_{\phi}s & = & \phi^{m}\left(\gamma_{m}\bar{\lambda}\right).
\end{eqnarray}
\end{subequations}For this reason, $\omega$, $\overline{\omega}$
and $s$ will always appear in the gauge invariant combinations:
\begin{equation}
\begin{array}{cc}
T_{\lambda}=-\omega\partial\lambda, & N^{mn}=-\frac{1}{2}\omega\gamma^{mn}\lambda,\\
T_{\bar{\lambda}}=-\bar{\omega}\partial\bar{\lambda}-s\partial r, & \overline{N}^{mn}=\frac{1}{2}\left(\bar{\lambda}\gamma^{mn}\bar{\omega}-r\gamma^{mn}s\right),\\
J_{\lambda}=-\omega\lambda, & J_{\bar{\lambda}}=-\bar{\lambda}\bar{\omega},\\
J_{r}=rs. & \Phi=r\bar{\omega},\\
S=\overline{\lambda}s, & S^{mn}=\frac{1}{2}\bar{\lambda}\gamma^{mn}s.
\end{array}\label{eq:gaugeinvariantobjects}
\end{equation}
All the relevant OPE's (for both matter and ghost sector) are presented
in the appendix.

Since worldsheet reparametrization symmetry is not explicit in \eqref{eq:matteraction},
\eqref{eq:actionminps} and \eqref{eq:actionnmps}, the theory lacks
the natural perturbative description that comes with the usual $\left(b,c\right)$
ghost system. Hence, the $b$ ghost in the pure spinor formalism,
as introduced in \cite{Berkovits:2004px,Berkovits:2005bt}, is built
out of the above variables to satisfy
\begin{equation}
\left\{ Q_{0},b\right\} =T,\label{eq:QbT}
\end{equation}
where $T$ is the total energy-momentum tensor. Its construction is
based on an ingenious chain of operators (geometrically described
in \cite{Berkovits:2006vi} and operationally reviewed in \cite{Oda:2007ak,Jusinskas:2013yca})
and the\emph{ }full quantum operator\emph{ }can be cast as
\begin{multline}
b=-s^{\alpha}\partial\overline{\lambda}_{\alpha}-\partial\left(\frac{\overline{\lambda}_{\alpha}\overline{\lambda}_{\beta}}{\left(\overline{\lambda}\lambda\right)^{2}}\right)\lambda^{\alpha}\partial\theta^{\beta}+\left(\frac{\overline{\lambda}_{\alpha}}{\left(\overline{\lambda}\lambda\right)},G^{\alpha}\right)-2!\left(\frac{\overline{\lambda}_{\alpha}r_{\beta}}{\left(\overline{\lambda}\lambda\right)^{2}},H^{\alpha\beta}\right)\\
-3!\left(\frac{\overline{\lambda}_{\alpha}r_{\beta}r_{\gamma}}{\left(\overline{\lambda}\lambda\right)^{3}},K^{\alpha\beta\gamma}\right)+4!\left(\frac{\overline{\lambda}_{\alpha}r_{\beta}r_{\gamma}r_{\lambda}}{\left(\overline{\lambda}\lambda\right)^{4}},L^{\alpha\beta\gamma\lambda}\right)\label{eq:quantumb}
\end{multline}
where\begin{subequations}
\begin{eqnarray}
G^{\alpha} & = & \frac{1}{2}\gamma_{m}^{\alpha\beta}\left(\Pi^{m},d_{\beta}\right)-\frac{1}{4}N_{mn}\left(\gamma^{mn}\partial\theta\right)^{\alpha}-\frac{1}{4}J_{\lambda}\partial\theta^{\alpha}+4\partial^{2}\theta^{\alpha},\label{eq:galpha}\\
H^{\alpha\beta} & = & \frac{1}{4\cdot96}\gamma_{mnp}^{\alpha\beta}\left(\frac{\alpha'}{2}d\gamma^{mnp}d+24N^{mn}\Pi^{p}\right),\\
K^{\alpha\beta\gamma} & = & -\frac{1}{96}\left(\frac{\alpha'}{2}\right)N_{mn}\gamma_{mnp}^{[\alpha\beta}\left(\gamma^{p}d\right)^{\gamma]},\\
L^{\alpha\beta\gamma\lambda} & = & -\frac{3}{\left(96\right)^{2}}\left(\frac{\alpha'}{2}\right)\left(N^{mn},N^{rs}\right)\eta^{pq}\gamma_{mnp}^{[\alpha\beta}\gamma_{qrs}^{\gamma]\lambda}.
\end{eqnarray}
\end{subequations}Observe that the ordering prescription is needed
here, for one is working with fields that diverge when approach each
other, and is implemented through:
\begin{equation}
\left(A,B\right)\left(y\right)\equiv\frac{1}{2\pi i}\oint\frac{dz}{z-y}A\left(z\right)B\left(y\right),\label{eq:ordering}
\end{equation}

The $b$ ghost in \eqref{eq:quantumb} is a conformal weight $2$
primary field,
\begin{equation}
T\left(z\right)b\left(y\right)\sim2\frac{b}{\left(z-y\right)^{2}}+\frac{\partial b}{\left(z-y\right)}.
\end{equation}
Note also that it is a Lorentz scalar, manifestly supersymmetric and
nilpotent \cite{Chandia:2010ix,Jusinskas:2013yca},
\begin{equation}
b\left(z\right)b\left(y\right)\sim\textrm{regular}.\label{eq:nilpotency}
\end{equation}

Together, $b$ and $J_{BRST}$ are the fermionic generators of a twisted
$\mathcal{N}=2$ $\hat{c}=3$ algebra \cite{Berkovits:2005bt},\begin{subequations}
\begin{eqnarray}
T\left(z\right)T\left(y\right) & \sim & 2\frac{T}{\left(z-y\right)^{2}}+\frac{\partial T}{\left(z-y\right)},\\
J_{BRST}\left(z\right)b\left(y\right) & \sim & \frac{3}{\left(z-y\right)^{3}}+\frac{J}{\left(z-y\right)^{2}}+\frac{T}{\left(z-y\right)},\label{eq:Jbrstb}\\
J\left(z\right)T\left(y\right) & \sim & \frac{3}{\left(z-y\right)^{3}}+\frac{J}{\left(z-y\right)^{2}},\label{eq:U1notprimary}\\
J\left(z\right)J\left(y\right) & \sim & \frac{3}{\left(z-y\right)^{2}},\label{eq:JJ}\\
J\left(z\right)J_{BRST}\left(y\right) & \sim & \frac{J_{BRST}}{\left(z-y\right)},\label{eq:U1chargeBRST}\\
J\left(z\right)b\left(y\right) & \sim & -\frac{b}{\left(z-y\right)},\label{eq:U1chargeb}
\end{eqnarray}
\end{subequations}where the bosonic generators are the energy-momentum
tensor, $T$, and the $U\left(1\right)$ current, $J$, given by
\begin{eqnarray}
J & = & J_{\lambda}+J_{r}-2\frac{\overline{\lambda}\partial\lambda}{\overline{\lambda}\lambda}+2\frac{r\partial\theta}{\overline{\lambda}\lambda}-2\frac{\left(r\lambda\right)\left(\overline{\lambda}\partial\theta\right)}{\left(\overline{\lambda}\lambda\right)^{2}}.
\end{eqnarray}
It is worth noting that, although unusual, the non-quadratic terms
appearing here ensure the closure of the algebra and do not spoil
the interpretation of $J$ as the ghost number current%
\footnote{Even the apparent mixing between matter and ghosts due to the non-regular
OPE
\[
d_{\alpha}\left(z\right)J\left(y\right)\sim\frac{2}{\left(z-y\right)^{2}}\left[Q_{0},\frac{\overline{\lambda}_{\alpha}}{\left(\overline{\lambda}\lambda\right)}\right]
\]
does not contradict this interpretation, as it is BRST-exact.%
}.

\section{Deforming the $b$ ghost\label{sec:deformations}}

\

From equation \eqref{eq:QbT}, it is clear that the $b$ ghost can
be defined only up to BRST-exact terms. In this sense, it is not unique
and it might be interesting to check whether the basic properties
presented in section \ref{sec:review} are preserved with a BRST-exact
deformed version of $b$.

For example, the simplest known version of the pure spinor $b$ ghost,
\begin{equation}
b_{nc}=-s\partial\overline{\lambda}+\left(\frac{C_{\alpha}}{C\lambda},G^{\alpha}\right)+2\frac{\left(C\partial\lambda\right)\left(C\partial\theta\right)}{\left(C\lambda\right)^{2}},\label{eq:bnoncov}
\end{equation}
differs from \eqref{eq:quantumb} by a BRST-exact term \cite{Oda:2007ak}.
It obviously satisfies $\left\{ Q_{0},b_{nc}\right\} =T$, but it
is non-covariant due to the presence of the constant spinor $C_{\alpha}$.

Performing the OPE computation of $b_{nc}$ with itself, given by
\begin{equation}
b_{nc}\left(z\right)b_{nc}\left(y\right)\sim\frac{1}{\left(z-y\right)}\frac{\left(C\gamma_{m}C\right)}{2\left(C\lambda\right)^{2}}\left\{ \Pi^{m}\underbrace{\left(-\frac{1}{\alpha'}\Pi^{2}-d\partial\theta\right)}_{A}-\frac{\alpha'}{8}\underbrace{\left(d\gamma^{m}\partial d\right)}_{D^{m}}+\ldots\right\} ,\label{eq:OPEbbnoncov}
\end{equation}
one readily observes that $b_{nc}$ is nilpotent%
\footnote{Observe that $A$, the Virasoro constraint, and $D^{m}$, belong to
Siegel's algebra \cite{Siegel:1985xj}, as expected, since $G^{\alpha}$
given in \eqref{eq:galpha} contains $\Pi^{m}\left(\gamma_{m}d\right)^{\alpha}$,
the $\kappa$-symmetry generator. As a possible pole in \eqref{eq:OPEbbnoncov}
should be BRST-closed \cite{Chandia:2010ix,Jusinskas:2013yca}, it
is straightforward to determine the remaining terms appearing inside
the curly brackets of \eqref{eq:OPEbbnoncov}, up to BRST-exact ones,
like $\left\{ Q_{0},\partial\Pi^{m}\frac{C\partial\theta}{C\lambda}\right\} $.%
} only for $\left(C\gamma_{m}C\right)=0$, that is, when $C_{\alpha}$
is a pure spinor.

Consider, now,
\begin{equation}
b'=b+\left[Q_{0},\beta\right].\label{eq:deformedb}
\end{equation}

Due to \eqref{eq:nilpotency}, it is clear that
\begin{equation}
b'\left(z\right)b'\left(y\right)\sim b\left(z\right)\left[Q_{0},\beta\left(y\right)\right]+\left[Q_{0},\beta\left(z\right)\right]b\left(y\right)+\left[Q_{0},\beta\left(z\right)\right]\left[Q_{0},\beta\left(y\right)\right].\label{eq:b'b'}
\end{equation}
Note that the left-hand side of this relation can be written as
\[
\left\{ Q_{0},\beta\left(z\right)b\left(y\right)-b\left(z\right)\beta\left(y\right)+\beta\left(z\right)\left[Q_{0},\beta\left(y\right)\right]\right\} +T\left(z\right)\beta\left(y\right)-\beta\left(z\right)T\left(y\right).
\]
Requiring $\beta$ to be a primary conformal weight $2$ object,
\begin{equation}
T\left(z\right)\beta\left(y\right)-\beta\left(z\right)T\left(y\right)\sim\textrm{regular},
\end{equation}
and equation \eqref{eq:b'b'} is equivalent to
\begin{equation}
b'\left(z\right)b'\left(y\right)\sim\left\{ Q_{0},\left(\beta\left(z\right)b\left(y\right)-b\left(z\right)\beta\left(y\right)+\beta\left(z\right)\left[Q_{0},\beta\left(y\right)\right]\right)\right\} .\label{eq:OPEb'b'}
\end{equation}
There is no hope that $b'$ will be nilpotent for a generic $\beta$,
as in the non-covariant example above, and to understand the general
case, it is useful to start with a simpler one.

\subsection{$\beta=\left(S,\frac{\overline{\lambda}\partial\theta}{\overline{\lambda}\lambda}\right)$\label{sub:beta-example}}

\

Consider the particular covariant deformation\begin{subequations}
\begin{eqnarray}
\beta & = & \left(S,\frac{\overline{\lambda}\partial\theta}{\overline{\lambda}\lambda}\right),\label{eq:samplephi}\\
\left[Q_{0},\beta\right] & = & -\left(\left(J_{\overline{\lambda}}+J_{r}\right),\frac{\overline{\lambda}\partial\theta}{\overline{\lambda}\lambda}\right)-\left(S,\frac{\overline{\lambda}\partial\lambda}{\overline{\lambda}\lambda}\right)\nonumber \\
 & + & \left(S,\frac{r\partial\theta}{\overline{\lambda}\lambda}\right)-\left(S,\frac{\left(\overline{\lambda}\partial\theta\right)\left(r\lambda\right)}{\left(\overline{\lambda}\lambda\right)^{2}}\right),
\end{eqnarray}
\end{subequations}where $S$ was defined in \eqref{eq:gaugeinvariantobjects}.

It is direct to see that
\begin{equation}
\beta\left(z\right)\left[Q_{0},\beta\left(y\right)\right]\sim\textrm{regular}.\label{eq:deformedOPE1}
\end{equation}
There are no double poles, and possible simple poles are proportional
to $\left(\overline{\lambda}\partial\theta\right)^{2}=0$.

The OPE between $\beta$ and $b$ is also simple to obtain. The action
of $S$ in $b$ is to transform $r_{\alpha}$ in $\overline{\lambda}_{\alpha}$,
making all the terms in the chain vanish due to the antisymmetric
form of $H^{\alpha\beta}$, $K^{\alpha\beta\gamma}$ and $L^{\alpha\beta\gamma\lambda}$.
There are simple poles proportional to $\left(\overline{\lambda}\partial\theta\right)^{2}$
and also quadratic poles related to the contraction between $d_{\alpha}$
and $\partial\theta^{\beta}$ (and, of course, simple poles coming
from the Taylor expansion), but they always appear together with the
constraints \eqref{eq:psconstraints}, implying that
\begin{equation}
\beta\left(z\right)b\left(y\right)\sim\textrm{regular}.\label{eq:deformedOPE2}
\end{equation}
Looking back to expression \eqref{eq:OPEb'b'} and using \eqref{eq:deformedOPE1}
and \eqref{eq:deformedOPE2}, nilpotency of the deformed $b$ ghost
\begin{equation}
b_{a}\equiv b+a\left[Q_{0},\left(S,\frac{\overline{\lambda}\partial\theta}{\overline{\lambda}\lambda}\right)\right],\label{eq:bghosta}
\end{equation}
follows directly, \emph{i.e.}
\begin{equation}
b_{a}\left(z\right)b_{a}\left(y\right)\sim\textrm{regular}.\label{eq:nilpotencydeformed}
\end{equation}
Here, $a$ is just a numerical constant.

As a final check, the OPE computation of $b_{a}$ with the BRST current
results
\begin{equation}
J_{BRST}\left(z\right)b_{a}\left(y\right)\sim\frac{3}{\left(z-y\right)^{3}}+\frac{J_{a}}{\left(z-y\right)^{2}}+\frac{T}{\left(z-y\right)},
\end{equation}
where
\begin{equation}
J_{a}=J_{\lambda}-aJ_{\overline{\lambda}}+\left(1-a\right)J_{r}+(8a-2)\left(\frac{\overline{\lambda}\partial\lambda}{\overline{\lambda}\lambda}-\frac{r\partial\theta}{\overline{\lambda}\lambda}+\frac{\left(r\lambda\right)\left(\overline{\lambda}\partial\theta\right)}{\left(\overline{\lambda}\lambda\right)^{2}}\right).\label{eq:deformedU1ex}
\end{equation}
Observe that
\begin{eqnarray*}
\oint dz\left\{ J_{a}\left(z\right)\frac{\overline{\lambda}_{\alpha}}{\left(\overline{\lambda}\lambda\right)}\left(y\right)\right\}  & = & -\frac{\overline{\lambda}_{\alpha}}{\left(\overline{\lambda}\lambda\right)}\left(y\right),\\
\oint dz\left\{ J_{a}\left(z\right)\frac{r_{\alpha}}{\left(\overline{\lambda}\lambda\right)}\left(y\right)\right\}  & = & 0,
\end{eqnarray*}
and the right-hand sides of the above equations do not depend on $a$.

Together, $b_{a}$, $J_{a}$, $J_{BRST}$ and $T$ satisfy a $\mathcal{N}=2$
$\hat{c}=3$ critical topological string algebra.

To illustrate how this particular example might be interesting, observe
that \eqref{eq:deformedU1ex} admits three simplifications, depending
on the numerical choices of $a$:
\begin{itemize}
\item the first one is trivial, $a=0$, and corresponds to the usual construction,
without deformations;
\item the second choice is $a=1$, removing $J_{r}$ from the $U\left(1\right)$
current. In this case, the combination $\left(J_{\lambda}-J_{\overline{\lambda}}\right)$
is explicit, but does not appear alone.
\item and the last one is $a=\frac{1}{4}$. With this particular choice,
\eqref{eq:deformedU1ex} is more conventional looking, since the unusual
non-quadratic-terms vanish:
\begin{equation}
J_{\frac{1}{4}}=J_{\lambda}-\frac{1}{4}J_{\overline{\lambda}}+\frac{3}{4}J_{r}.\label{eq:U1conventional}
\end{equation}
Note that the non-minimal variables become fractionally charged.
\end{itemize}

\subsection{The invariance of the topological string algebra}

\

Now, a more general class of deformations (defined as in \eqref{eq:deformedb})
will be analysed.

Requiring invariance of the $\mathcal{N}=2$ $\hat{c}=3$ algebra,
it will be shown that some constraints on the deformations must be
imposed and $\beta$ will be restricted to be:
\begin{itemize}
\item a commuting object, in order for $b'$ to have definite statistics;
\item an ghost number $-2$ object with respect to $J$ (then $b'$ will
have a definite ghost charge), that is
\begin{equation}
J\left(z\right)\beta\left(y\right)\sim-2\frac{\beta}{\left(z-y\right)}.\label{eq:U1chargedeformation}
\end{equation}

\item supersymmetric, which avoids the explicit introduction of objects
that trivialize the cohomology. The two known examples are\begin{subequations}
\begin{eqnarray}
\xi_{1} & = & \frac{\left(C\theta\right)}{\left(C\lambda\right)}, \label{eq:noncovtriv}\\
\xi_{2} & = & \frac{\left(\overline{\lambda}\theta\right)}{\left(\overline{\lambda}\lambda\right)-\left(r\theta\right)},
\end{eqnarray}
\end{subequations}where $\left\{ Q_{0},\xi_{1}\right\} =\left\{ Q_{0},\xi_{2}\right\} =1$
and $C_{\alpha}$ is any constant spinor \footnote{The constructions with constant spinors in \eqref{eq:bnoncov} and
\eqref{eq:noncovtriv} are a bit subtle, since they not globally defined
in the pure spinor space. More details can be found in \cite{Bedoya:2011kv}.};
\item and, as already mentioned, a primary conformal weight $2$ field,
\begin{equation}
T\left(z\right)\beta\left(y\right)\sim2\frac{\beta}{\left(z-y\right)^{2}}+\frac{\partial\beta}{\left(z-y\right)};\label{eq:primarydeformation}
\end{equation}

\end{itemize}
\

With this in mind, it is straightforward to examine the impact of
the deformation on the topological string algebra.

Assuming that the BRST current does not change, the first relation
that will be presented is the OPE between $J_{BRST}$ and $\beta$,
that can be generically written as
\begin{equation}
J_{BRST}\left(z\right)\beta\left(y\right)\sim\frac{\left[Q_{1},\beta\right]-y\left[Q_{0},\beta\right]}{\left(z-y\right)^{2}}+\frac{\left[Q_{0},\beta\right]}{\left(z-y\right)},\label{eq:BRSTbeta}
\end{equation}
where $Q_{n}$ was defined in \eqref{eq:BRSTmodes}. Note that the
cubic pole vanishes, since there are no ghost number $-1$ anticommuting
world-sheet scalars with the above requisites (for example, $\left(\overline{\lambda}\theta\right)$
is ruled out as it is not supersymmetric). Then, it follows that
\begin{equation}
J_{BRST}\left(z\right)\left[Q_{0},\beta\right]\left(y\right)\sim\frac{J'-J}{\left(z-y\right)^{2}},
\end{equation}
for the BRST charge $Q_{0}$ is nilpotent. The quadratic pole does
not have to vanish and the deformed $U\left(1\right)$ current is
defined to be
\begin{equation}
J'=J-\left\{ Q_{0},\left[Q_{1},\beta\right]\right\} \label{eq:deformedU1}
\end{equation}
Therefore, \eqref{eq:Jbrstb} is reproduced with $J\to J'$ and $b\to b'$.

The next OPE, \eqref{eq:U1notprimary}, is obviously preserved, since
$\beta$ is a primary conformal weight $2$ field by assumption.

The $J'$ OPE with itself is given generically by
\begin{equation}
J'\left(z\right)J'\left(y\right)\sim\left\{ \frac{3+\varphi}{\left(z-y\right)^{2}}+\frac{\frac{1}{2}\partial\varphi}{\left(z-y\right)}\right\} ,\label{eq:J'J'}
\end{equation}
where the contribution of $J\left(z\right)J\left(y\right)$ was made
explicit. Observe that,
\begin{eqnarray*}
J\left(z\right)\left\{ Q_{0},\left[Q_{1},\beta\left(y\right)\right]\right\}  & = & \left\{ Q_{0},J\left(z\right)\left[Q_{1},\beta\left(y\right)\right]\right\} +J_{BRST}\left(z\right)\left[Q_{1},\beta\left(y\right)\right]\\
 & = & -\left\{ Q_{0},\left[Q_{1},J\left(z\right)\right]\beta\left(y\right)\right\} -J_{BRST}\left(z\right)\left[Q_{1},\beta\right]\left(y\right)\\
 &  & +\left\{ Q_{0},\left[Q_{1},J\left(z\right)\beta\left(y\right)\right]\right\} .
\end{eqnarray*}
As $\left[Q_{1},J\left(z\right)\right]=-zJ_{BRST}\left(z\right)$,
the right-hand side can be rewritten as
\begin{eqnarray}
J\left(z\right)\left\{ Q_{0},\left[Q_{1},\beta\left(y\right)\right]\right\}  & = & \left\{ Q_{0},\left[Q_{1},J\left(z\right)\beta\left(y\right)\right]\right\} \nonumber \\
 &  & -z\left\{ Q_{0},J_{BRST}\left(z\right)\beta\left(y\right)\right\} +\left\{ Q_{1},J_{BRST}\left(z\right)\beta\left(y\right)\right\} .
\end{eqnarray}
Noting that
\begin{equation}
\left\{ Q_{n},Q_{m}\right\} =0,\label{eq:antibrstmodes}
\end{equation}
for any $n,m\geq0$, equation \eqref{eq:BRSTbeta} implies that $J\left(z\right)\left\{ Q_{0},\left[Q_{1},\beta\left(y\right)\right]\right\} $
is BRST-exact. Thus, replacing the definition \eqref{eq:deformedU1}
in left hand side of \eqref{eq:J'J'}, $\varphi$ is demonstrated
to be a ghost number $0$ BRST-exact world-sheet scalar, which cannot
appear due to the hypothesis on $\beta$ and shows that the OPE \eqref{eq:JJ}
is reproduced when $J\to J'$.

Going on, the OPE with $J_{BRST}$ and $\left(J'-J\right)$ can be
trivially shown to be regular, as its general form can be cast as
\begin{equation}
J_{BRST}\left(z\right)\left\{ Q_{1},\left[Q_{0},\beta\right]\right\} \left(0\right)\sim\frac{\left[Q_{1}\left\{ Q_{1},\left[Q_{0},\beta\right]\right\} \right]}{z^{2}}+\frac{\left[Q_{0},\left\{ Q_{1},\left[Q_{0},\beta\right]\right\} \right]}{z}.
\end{equation}
Through \eqref{eq:antibrstmodes}, this result demonstrates that the
deformations preserve \eqref{eq:U1chargeBRST}.

The last OPE to be analysed is \eqref{eq:U1chargeb}, with $J\to J'$
and $b\to b'$.
\begin{eqnarray}
J'\left(z\right)b'\left(y\right) & = & J\left(z\right)b\left(y\right)+\left\{ Q_{1},\left[Q_{0},\beta\left(z\right)\right]\right\} b\left(y\right)\nonumber \\
 &  & +J\left(z\right)\left[Q_{0},\beta\left(y\right)\right]+\left\{ Q_{1},\left[Q_{0},\beta\left(z\right)\right]\right\} \left[Q_{0},\beta\left(y\right)\right].
\end{eqnarray}
Using \eqref{eq:U1chargedeformation}, \eqref{eq:BRSTbeta} and $\left\{ Q_{1},b\left(z\right)\right\} =J\left(z\right)+zT\left(z\right)$,
the above equation can be rewritten as
\begin{eqnarray}
J'\left(z\right)b'\left(y\right) & \sim & -\frac{b'}{\left(z-y\right)}\nonumber \\
 &  & +\left[Q_{0},\left\{ Q_{1},\left[Q_{0},\beta\left(z\right)\right]\right\} \beta\left(y\right)\right]\nonumber \\
 &  & -\left[Q_{0},\left\{ Q_{1},\beta\left(z\right)b\left(y\right)\right\} \right]\label{eq:deformedbU1charge}
\end{eqnarray}

In order for the topological algebra to be preserved, equations \eqref{eq:OPEb'b'}
and \eqref{eq:deformedbU1charge} impose some conditions on $\beta$
and the following OPE's must hold up to BRST-closed poles:\begin{subequations}\label{eq:betaconstraints}
\begin{eqnarray}
\beta\left(z\right)b\left(y\right) & \sim & \textrm{regular},\label{eq:betabclosed}\\
\beta\left(z\right)\left[Q_{0},\beta\left(y\right)\right] & \sim & \textrm{regular},\\
\beta\left(z\right)\left\{ Q_{0},\left[Q_{1},\beta\left(y\right)\right]\right\}  & \sim & \textrm{regular}.
\end{eqnarray}
\end{subequations}If this is the case,
\begin{eqnarray*}
b'\left(z\right)b'\left(y\right) & \sim & \textrm{regular},\\
J'\left(z\right)b'\left(y\right) & \sim & -\frac{b'}{\left(z-y\right)}.
\end{eqnarray*}

Therefore, the $\mathcal{N}=2$ $\hat{c}=3$ critical topological
string algebra is invariant under the self-consistent deformations
of the $b$ ghost and the $U\left(1\right)$ current, respectively,
\eqref{eq:deformedb} and \eqref{eq:deformedU1}, as long as the requisites
on $\beta$ presented before equation \eqref{eq:BRSTbeta} and in
\eqref{eq:betaconstraints} are imposed. The example of subsection
\ref{sub:beta-example} satisfies all of these conditions.

Note that equation \eqref{eq:betabclosed} opens the question about
the cohomology of the $b$ ghost. It is interesting to point out that
in the same manner that the BRST cohomology is non-trivial only for
world-sheet scalars, the $b$ ghost cohomology can be shown to be
non-trivial only under a certain condition, that will now be derived.

Defining,
\begin{equation}
B\equiv\oint dzb\left(z\right),
\end{equation}
it is direct to demonstrate through the OPE \eqref{eq:Jbrstb} that
\begin{equation}
\left\{ B,J_{BRST}\left(z\right)\right\} =T\left(z\right)-\partial J\left(z\right).
\end{equation}

Now, suppose that there is an operator $V_{hg}$ satisfying\begin{subequations}
\begin{eqnarray}
T\left(z\right)V_{hg}\left(y\right) & \sim & h\frac{V_{hg}}{\left(z-y\right)^{2}}+\frac{\partial V_{hg}}{\left(z-y\right)},\\
J\left(z\right)V_{hg}\left(y\right) & \sim & g\frac{V_{hg}}{\left(z-y\right)},
\end{eqnarray}
\end{subequations}and that is annihilated by $B$, \emph{i.e.}
\begin{equation}
\left[B,V_{hg}\right]=0.
\end{equation}
Then it follows that
\begin{eqnarray}
\left\{ B,J_{BRST}\left(z\right)V_{hg}\left(y\right)\right\}  & \sim & \left(h+g\right)\frac{V_{hg}}{\left(z-y\right)^{2}}+\left(1-g\right)\frac{\partial V_{hg}}{\left(z-y\right)},
\end{eqnarray}
showing that $V_{hg}$ is $B$-exact for $\left(h+g\right)\neq0$
and constituting an exclusion\emph{ }criterion\emph{ }for the non-trivial
cohomology of $B$.

As can be seen by equations \eqref{eq:U1chargedeformation} and \eqref{eq:primarydeformation},
$\beta$ in the $b$ ghost deformations discussed above satisfies
$\left(h+g\right)=0$.

The cohomology of $B$ will not be further discussed. Note that even
the space where $B$ acts is not yet understood. For example, one
has to be concerned about poles em $\left(\overline{\lambda}\lambda\right)$
higher than $11$, as there is not a simple regularization scheme
that would allow a formal functional integration over its zero modes
\cite{Berkovits:2006vi}. Note also that there is no natural candidate
for an operator that trivializes the $B$ cohomology. Next section
discusses a possible construction that fits the desired properties
of the $c$ ghost.

\section{A possible $c$ ghost\label{sec:cghost}}

\

In the topological string perspective, the existence of a $c$ ghost
in the non-minimal pure spinor formalism may seem to be meaningless.
Indeed, the construction of the $b$ ghost conjugate is very unusual
and, more than that, unrequired. The reason is simple. First, one
does not have a natural $-1$ conformal weight field to work with.
Second, the amplitudes prescription (including the notion of unintegrated
vertex, compared to the other superstring formalisms) is very well
established without it.

Under these conditions, a $c$ ghost like field is undoubtedly strange.
It will be defined as
\begin{equation}
c\equiv-\frac{\left(r\lambda\right)}{\left(\partial\overline{\lambda}\lambda\right)+\left(r\partial\theta\right)},\label{eq:cghost}
\end{equation}
satisfying the relation
\begin{equation}
b\left(z\right)c\left(y\right)\sim\frac{1}{\left(z-y\right)}.\label{eq:bcconjugation}
\end{equation}
Note that $c$ must have ghost number $+1$, since $b$ is a ghost
number $-1$ field.

To demonstrate that \eqref{eq:cghost} is the conjugate of \eqref{eq:quantumb},
observe that
\begin{equation}
\left(r\lambda\right)=-\left[Q_{0},\left(\overline{\lambda}\lambda\right)\right].
\end{equation}
By a direct computation, one can derive
\begin{equation}
b\left(z\right)\left(r\lambda\right)\left(y\right)\sim-\frac{\left(\partial\overline{\lambda}\lambda\right)+\left(r\partial\theta\right)}{\left(z-y\right)},
\end{equation}
which is verified through
\begin{eqnarray}
b\left(z\right)\left[Q_{0},\left(\overline{\lambda}\lambda\right)\right]\left(y\right) & = & -\left\{ Q_{0},b\left(z\right)\left(\overline{\lambda}\lambda\right)\left(y\right)\right\} +\left\{ Q_{0},b\left(z\right)\right\} \left(\overline{\lambda}\lambda\right)\left(y\right)\nonumber \\
 & = & -\left\{ Q_{0},b\left(z\right)\left(\overline{\lambda}\lambda\right)\left(y\right)\right\} +T\left(z\right)\left(\overline{\lambda}\lambda\right)\left(y\right)\nonumber \\
 & \sim & \frac{\partial\left(\overline{\lambda}\lambda\right)}{\left(z-y\right)}-\frac{\left\{ Q_{0},\left(\overline{\lambda}\partial\theta\right)\right\} }{\left(z-y\right)}\nonumber \\
 & \sim & \frac{\left(\partial\overline{\lambda}\lambda\right)+\left(r\partial\theta\right)}{\left(z-y\right)},
\end{eqnarray}
where \eqref{eq:QbT} was used.

Since the $b$ ghost is nilpotent, \eqref{eq:nilpotency}, the right
hand side of the above equation does not have any poles with $b$,
that is
\begin{equation}
b\left(z\right)\left(\partial\overline{\lambda}\lambda+r\partial\theta\right)\left(y\right)\sim\textrm{regular}.
\end{equation}
Therefore, equation \eqref{eq:bcconjugation} is directly reproduced.

Note also that
\begin{equation}
\left\{ Q_{0},c\right\} =c\partial c,
\end{equation}
the usual BRST relation between the $c$ ghost and the BRST charge,
and that $c$ is a supersymmetric Lorentz scalar.

The analogous construction of the $c$ ghost as the conjugate of \eqref{eq:bghosta}
is
\begin{equation}
c_{a}\equiv-\frac{\left(r\lambda\right)}{\partial\left(\overline{\lambda}\lambda\right)+\left(a-1\right)\left\{ Q_{0},\left(\overline{\lambda}\partial\theta\right)\right\} },
\end{equation}
satisfying
\begin{equation}
b_{a}\left(z\right)c_{a}\left(y\right)\sim\frac{1}{\left(z-y\right)}.
\end{equation}

\section{Summary and conclusions}

\

In this work, a simple analysis of the non-uniqueness on the definition
of the $b$ ghost in the non-minimal pure spinor formalism was made.

As the relation $\left\{ Q_{0},b\right\} =T$ allows the construction
of the composite $b$ ghost up to BRST-exact terms, it is interesting
to understand the emerging of the critical topological string algebra
with the different versions of such field. Observe that for
\begin{equation}
\delta b=\left[Q_{0},\beta\right],
\end{equation}
self-consistency in the $\mathcal{N}=2$ algebra implies a deformation
of the $U\left(1\right)$ current as well,
\begin{equation}
\delta J=\left\{ Q_{1},\left[Q_{0},\beta\right]\right\} .
\end{equation}
For a particular example, it was shown to appear in a more conventional
form (equation \eqref{eq:U1conventional}), and nilpotency of the
deformed $b$ ghost was demonstrated (equation \eqref{eq:nilpotencydeformed}).

In section \ref{sec:cghost}, a novel feature of the formalism was
introduced, the $c$ ghost. Although interesting, the strange form
of \eqref{eq:cghost} may be pathological, in the sense that one now
is able to construct an entire new class of composite operators that
trivialize the cohomology, \emph{e.g.}
\begin{equation}
\xi\equiv\frac{\overline{\lambda}\partial\theta}{\overline{\lambda}\partial\lambda-r\partial\theta}\Rightarrow\left\{ Q_{0},\xi\right\} =1.
\end{equation}
It is clear, however, that this construction is highly artificial
and cannot emerge naturally in any known process for the pure spinor
formalism. From the conformal field theory point of view, this kind
of construction is very unusual. Note that the denominator in \eqref{eq:cghost}
contains derivatives of world-sheet scalars, which implies that, wherever
they vanish, the $c$ ghost is singular.

Note also that the existence of a composite field satisfying
\[
b\left(z\right)c\left(y\right)\sim\frac{1}{\left(z-y\right)},
\]
trivializes the cohomology of $b$. In the twisting picture, the BRST
current $J_{BRST}$ and the $b$ ghost exchange roles in different
twists. Then, it might be useful to understand the cohomology of the
pure spinor $b$ ghost and study the Siegel's gauge implementation
on the physical vertices (\emph{e.g. }\cite{Aisaka:2009yp}).

The simple results shown here evidence that the pure spinor $b$ ghost
is only partially understood. A deeper example, is the recent work
of N. Berkovits \cite{Berkovits:2013pla}, which presents the structure
\begin{equation}
\overline{\Gamma}^{m}=\frac{1}{2}\frac{\left(\overline{\lambda}\gamma^{m}d\right)}{\left(\overline{\lambda}\lambda\right)}-\frac{1}{8}\frac{\left(\overline{\lambda}\gamma^{mnp}r\right)N_{np}}{\left(\overline{\lambda}\lambda\right)^{2}}.\label{eq:gammabar}
\end{equation}
More than the visual simplification of the $b$ ghost, which can now
be written as
\[
b=\Pi_{m}\overline{\Gamma}^{m}+\frac{1}{4}\frac{\left(r\gamma_{mn}\lambda\right)}{\left(\overline{\lambda}\lambda\right)}\overline{\Gamma}^{m}\overline{\Gamma}^{n}+\ldots,
\]
the introduction of \eqref{eq:gammabar} is the first step of an attempt
to relate the pure spinor formalism to the RNS string in which was
called a dynamical twisting.

All together, these newly found ingredients may lead to a better understanding
of the world-sheet origin of the pure spinor formalism and be the
first step, perhaps, in establishing an equivalence with the other
known superstring formalisms%
\footnote{In a recent work \cite{Donagi:2013dua}, it was shown that the super
\emph{moduli} space is not projected for genus $g\geq5$, meaning
that ``certain approaches to superstring perturbation theory that
are very powerful in low orders have no close analogue in higher orders''.
Basically, the super \emph{moduli} space has a structure on its own,
which implies that the pure spinor formalism may not be equivalent
to the RNS superstring as it does not contain such structure. This
deserves further investigation.%
}.

\

\textbf{Acknowledgements:} I would like to thank Ido Adam, Nathan
Berkovits, Humberto Gomez, Sebastian Guttenberg and Andrei Mikhailov
for useful discussions and suggestions. This work was supported by
FAPESP grant 2009/17516-4.

\appendix

\section{Appendix}

\subsection{Conventions}

\

Indices:
\[
\begin{cases}
m,n,\ldots=0,\ldots,9 & \textrm{space-time vector indices,}\\
\alpha,\beta,\ldots=1,\ldots,16 & \textrm{space-time spinor indices,}
\end{cases}
\]

The indices antisymmetrization is represented by the square brackets,
meaning
\begin{equation}
\left[I_{1}\ldots I_{n}\right]=\frac{1}{n!}\left(I_{1}\ldots I_{n}+\textrm{all antisymmetric permutations}\right).
\end{equation}
For example,
\begin{equation}
\gamma^{[m}\gamma^{n]}=\frac{1}{2}\left(\gamma^{m}\gamma^{n}-\gamma^{n}\gamma^{m}\right)=\gamma^{mn}.
\end{equation}

Concerning OPE's, the right-hand sides of the equations are always
evaluated at the coordinate of the second entry, that is,
\begin{equation}
A\left(z\right)B\left(y\right)\sim\frac{C}{\left(z-y\right)^{2}}+\frac{D}{\left(z-y\right)}
\end{equation}
means $C=C\left(y\right)$ and $D=D\left(y\right)$.

The gamma matrices $\gamma_{\alpha\beta}^{m}$ and $\gamma_{m}^{\alpha\beta}$
satisfy
\begin{equation}
\left\{ \gamma^{m},\gamma^{n}\right\} _{\phantom{\alpha}\beta}^{\alpha}=\left(\gamma^{m}\right)^{\alpha\sigma}\gamma_{\sigma\beta}^{n}+\left(\gamma^{n}\right)^{\alpha\sigma}\gamma_{\sigma\beta}^{m}=2\eta^{mn}\delta_{\beta}^{\alpha}.\label{eq:diracalgebra}
\end{equation}

The Fierz decompositions of bispinors are given by
\begin{equation}
\chi^{\alpha}\psi^{\beta}=\frac{1}{16}\gamma_{m}^{\alpha\beta}\left(\chi\gamma^{m}\psi\right)+\frac{1}{3!16}\gamma_{mnp}^{\alpha\beta}\left(\chi\gamma^{mnp}\psi\right)+\frac{1}{5!16}\left(\frac{1}{2}\right)\gamma_{mnpqr}^{\alpha\beta}\left(\chi\gamma^{mnpqr}\psi\right)
\end{equation}
and
\begin{equation}
\chi_{\alpha}\psi^{\beta}=\frac{1}{16}\delta_{\alpha}^{\beta}\left(\chi\psi\right)-\frac{1}{2!16}\left(\gamma_{mn}\right)_{\phantom{\beta}\alpha}^{\beta}\left(\chi\gamma^{mn}\psi\right)+\frac{1}{4!16}\left(\gamma_{mnpq}\right)_{\phantom{\beta}\alpha}^{\beta}\left(\chi\gamma^{mnpq}\psi\right),
\end{equation}
which can be used to derive the following identities:\begin{subequations}
\begin{eqnarray}
\left(\gamma^{mn}\right)_{\phantom{\alpha}\beta}^{\alpha}\left(\gamma_{mn}\right)_{\phantom{\gamma}\lambda}^{\gamma} & = & 4\gamma_{\beta\lambda}^{m}\gamma_{m}^{\alpha\gamma}-2\delta_{\beta}^{\alpha}\delta_{\lambda}^{\gamma}-8\delta_{\lambda}^{\alpha}\delta_{\beta}^{\gamma},\\
\eta_{mn}\gamma_{\alpha\beta}^{m}\gamma_{\gamma\lambda}^{n} & = & -\eta_{mn}\left(\gamma_{\alpha\gamma}^{m}\gamma_{\beta\lambda}^{n}+\gamma_{\alpha\lambda}^{m}\gamma_{\gamma\beta}^{n}\right),\\
\gamma^{m}\gamma^{n_{1}\ldots n_{k}}\gamma_{m} & = & \left(-1\right)^{k}\left(10-2k\right)\gamma^{n_{1}\ldots n_{k}}.
\end{eqnarray}
\end{subequations}

\subsection{Fundamental OPE's\label{sub:fundamentalopes}}

\

This part of the Appendix is a summary of the relevant OPE's that
are being used in this work.

\subsubsection*{Matter fields}

\

The fundamental fields of the matter sector are $X^{m}\left(z,\bar{z}\right)$,
$p_{\alpha}\left(z\right)$ and $\theta^{\alpha}\left(z\right)$,
satisfying
\begin{eqnarray}
X^{m}\left(z,\bar{z}\right)X^{n}\left(y,\bar{y}\right) & \sim & -\frac{\alpha'}{2}\eta^{mn}\ln\left|z-y\right|^{2},\\
p_{\alpha}\left(z\right)\theta^{\beta}\left(y\right) & \sim & \frac{\delta_{\alpha}^{\beta}}{z-y}.
\end{eqnarray}

The OPE's involving the primary supersymmetric objects defined in
\eqref{eq:susyderivative} and \eqref{eq:susymomentum} are given
by\begin{subequations}
\begin{eqnarray}
\Pi^{m}\left(z\right)\Pi^{n}\left(y\right) & \sim & -\frac{\alpha'}{2}\frac{\eta^{mn}}{\left(z-y\right)^{2}},\\
d_{\alpha}\left(z\right)\Pi^{m}\left(y\right) & \sim & \frac{\gamma_{\alpha\beta}^{m}\partial\theta^{\beta}}{\left(z-y\right)},\\
d_{\alpha}\left(z\right)d_{\beta}\left(y\right) & \sim & -\frac{2}{\alpha'}\frac{\gamma_{\alpha\beta}^{m}\Pi_{m}}{\left(z-y\right)}.
\end{eqnarray}
\end{subequations}The matter energy momentum tensor is simply
\begin{equation}
T_{\textrm{matter}}=-\frac{1}{\alpha'}\Pi^{m}\Pi_{m}-d_{\alpha}\partial\theta^{a}
\end{equation}
and satisfies
\begin{equation}
T_{\textrm{matter}}\left(z\right)T_{\textrm{matter}}\left(y\right)\sim-\frac{11}{\left(z-y\right)^{4}}+2\frac{T_{\textrm{matter}}}{\left(z-y\right)^{2}}+\frac{\partial T_{\textrm{matter}}}{\left(z-y\right)}.
\end{equation}

\subsubsection*{Minimal fields}

\

As mentioned in section \ref{sec:review}, only gauge invariant quantities
are being used. The set of OPE's from the minimal sector is:
\[
\begin{array}{ccc}
T_{\lambda}\left(z\right)T_{\lambda}\left(y\right)\sim\frac{11}{\left(z-y\right)^{4}}+2\frac{T_{\lambda}}{\left(z-y\right)^{2}}+\frac{\partial T_{\lambda}}{\left(z-y\right)}, &  & J_{\lambda}\left(z\right)T_{\lambda}\left(y\right)\sim-\frac{8}{\left(z-y\right)^{3}}+\frac{J_{\lambda}}{\left(z-y\right)^{2}}\end{array}
\]
\[
\begin{array}{ccccc}
N^{mn}\left(z\right)T_{\lambda}\left(y\right)\sim\frac{N^{mn}}{\left(z-y\right)^{2}}, &  & T_{\lambda}\left(z\right)\lambda^{\alpha}\left(y\right)\sim\frac{\partial\lambda^{\alpha}}{\left(z-y\right)}, &  & N^{mn}\left(z\right)\lambda^{\alpha}\left(y\right)\sim\frac{1}{2}\frac{\left(\gamma^{mn}\lambda\right)^{\alpha}}{\left(z-y\right)},\end{array}
\]
\[
\begin{array}{ccccc}
N^{mn}\left(z\right)J_{\lambda}\left(y\right)\sim\textrm{regular}, &  & J_{\lambda}\left(z\right)\lambda^{\alpha}\left(y\right)\sim\frac{\lambda^{\alpha}}{\left(z-y\right)}, &  & J_{\lambda}\left(z\right)J_{\lambda}\left(y\right)\sim-\frac{4}{\left(z-y\right)^{2}},\end{array}
\]
\[
N^{mn}\left(z\right)N^{pq}\left(y\right)\sim6\frac{\eta^{m[p}\eta^{q]n}}{\left(z-y\right)^{2}}+2\frac{\eta^{m[q}N^{p]n}+\eta^{n[p}N^{q]m}}{\left(z-y\right)}.
\]

\subsubsection*{Non-minimal fields}

\

For the non-minimal sector, that encompasses a larger number of gauge
invariant currents, the full set of OPE's is:
\[
\begin{array}{ccc}
T_{\overline{\lambda}}\left(z\right)T_{\overline{\lambda}}\left(y\right)\sim2\frac{T_{\overline{\lambda}}}{\left(z-y\right)^{2}}+\frac{\partial T_{\overline{\lambda}}}{\left(z-y\right)}, & \overline{N}^{mn}\left(z\right)T_{\overline{\lambda}}\left(y\right)\sim\frac{\overline{N}^{mn}}{\left(z-y\right)^{2}}, & S^{mn}\left(z\right)T_{\overline{\lambda}}\left(y\right)\sim\frac{S^{mn}}{\left(z-y\right)^{2}},\end{array}
\]
\[
\begin{array}{ccc}
J_{\bar{\lambda}}\left(z\right)T_{\overline{\lambda}}\left(y\right)\sim-\frac{11}{\left(z-y\right)^{3}}+\frac{\overline{J}_{\overline{\lambda}}}{\left(z-y\right)^{2}}, & \Phi\left(z\right)T_{\overline{\lambda}}\left(y\right)\sim\frac{\Phi}{\left(z-y\right)^{2}}, & S\left(z\right)T_{\overline{\lambda}}\left(y\right)\sim\frac{S}{\left(z-y\right)^{2}},\end{array}
\]
\[
\begin{array}{ccc}
J_{r}\left(z\right)T_{\overline{\lambda}}\left(y\right)\sim\frac{11}{\left(z-y\right)^{3}}+\frac{J_{r}}{\left(z-y\right)^{2}}, & T_{\overline{\lambda}}\left(z\right)\overline{\lambda}_{\alpha}\left(y\right)\sim\frac{\partial\overline{\lambda}_{\alpha}}{\left(z-y\right)}, & T_{\overline{\lambda}}\left(z\right)r_{\alpha}\left(y\right)\sim\frac{\partial r_{\alpha}}{\left(z-y\right)},\end{array}
\]
\[
\begin{array}{ccc}
\Phi\left(z\right)S\left(y\right)\sim-\frac{8}{\left(z-y\right)^{2}}-\frac{J_{\overline{\lambda}}+J_{r}}{\left(z-y\right)}, & \Phi\left(z\right)S^{mn}\left(y\right)\sim\frac{\overline{N}^{mn}}{\left(z-y\right)}, & \Phi\left(z\right)\overline{\lambda}_{\alpha}\left(y\right)\sim-\frac{r_{\alpha}}{\left(z-y\right)},\end{array}
\]
\[
\begin{array}{ccc}
\Phi\left(z\right)\Phi\left(y\right)\sim\textrm{regular}, & \overline{N}^{mn}\left(z\right)J_{\bar{\lambda}}\left(y\right)\sim\textrm{regular}, & \overline{N}^{mn}\left(z\right)\Phi\left(y\right)\sim\textrm{regular},\end{array}
\]
\[
\begin{array}{ccc}
\overline{N}^{mn}\left(z\right)\overline{N}^{pq}\left(y\right)\sim2\frac{\eta^{m[q}\overline{N}^{p]n}+\eta^{n[p}\overline{N}^{q]m}}{\left(z-y\right)}, & J_{\bar{\lambda}}\left(z\right)J_{r}\left(y\right)\sim-\frac{3}{\left(z-y\right)^{2}}, & J_{\bar{\lambda}}\left(z\right)J_{\bar{\lambda}}\left(y\right)\sim-\frac{5}{\left(z-y\right)^{2}},\end{array}
\]
\[
\begin{array}{ccc}
\overline{N}^{mn}\left(z\right)J_{r}\left(y\right)\sim\textrm{regular}, & \overline{N}^{mn}\left(z\right)S\left(y\right)\sim\textrm{regular}, & J_{r}\left(z\right)J_{r}\left(y\right)\sim\frac{11}{\left(z-y\right)^{2}},\end{array}
\]
\[
\begin{array}{ccc}
\overline{N}^{mn}\left(z\right)\overline{\lambda}_{\alpha}\left(y\right)\sim-\frac{1}{2}\frac{\left(\overline{\lambda}\gamma^{mn}\right)_{\alpha}}{\left(z-y\right)}, & \overline{N}^{mn}\left(z\right)r_{\alpha}\left(y\right)\sim-\frac{1}{2}\frac{\left(r\gamma^{mn}\right)_{\alpha}}{\left(z-y\right)}, & J_{\bar{\lambda}}\left(z\right)\overline{\lambda}_{\alpha}\left(y\right)\sim\frac{\overline{\lambda}_{\alpha}}{\left(z-y\right)},\end{array}
\]
\[
\begin{array}{ccc}
J_{r}\left(z\right)r_{\alpha}\left(y\right)\sim\frac{r_{\alpha}\left(y\right)}{\left(z-y\right)}, & J_{\bar{\lambda}}\left(z\right)r_{\alpha}\left(y\right)\sim\textrm{regular}, & J_{r}\left(z\right)\overline{\lambda}_{\alpha}\left(y\right)\sim\textrm{regular}.\end{array}
\]

\pagebreak{}
\end{document}